**Low-energy Electron Reflectivity from Graphene**

R. M. Feenstra, N. Srivastava, Qin Gao, and M. Widom
Dept. Physics, Carnegie Mellon University, Pittsburgh, Pennsylvania 15213

Bogdan Diaconescu, Taisuke Ohta, and G. L. Kellogg
Sandia National Laboratories, Albuquerque, New Mexico 87185

J. T. Robinson
Naval Research Laboratory, Washington, D.C. 20375

I. V. Vlassiouk
Oak Ridge National Laboratory, P.O. Box 2008, Oak Ridge, Tennessee 37831

**Abstract**
Low-energy reflectivity of electrons from single- and multi-layer graphene is examined both theoretically and experimentally. A series of minima in the reflectivity over the energy range of 0 – 8 eV are found, with the number of minima depending on the number of graphene layers. Using first-principles computations, it is demonstrated that a free standing $n$-layer graphene slab produces $n-1$ reflectivity minima. This same result is also found experimentally for graphene supported on $SiO_2$. For graphene bonded onto other substrates it is argued that a similar series of reflectivity minima is expected, although in certain cases an additional minimum occurs, at an energy that depends on the graphene-substrate separation and the effective potential in that space.

The reflectivity of low-energy electrons from single- and multi-layer graphene has proven to be a very useful probe of the material. When examined over the energy range of about 0 – 8 eV, such spectra reveal a series of local maxima and minima. The minima in particular are important, since they reveal transmission maxima, i.e. *transmission resonances*, for the graphene. It was demonstrated in 2008 by Hibino et al. that, for *n* layers of graphene on a SiC(0001) surface, there are *n* minima in the spectra.[1] This relationship has provided the basis for subsequent works in which the thickness variation of the graphene on SiC is mapped out over the SiC wafer.[2,3,4,5,6] Hibino and co-workers presented a simple tight-binding model in which the transmission resonances arise from states localized on each graphene layer, with reflectivity minima formed by linear combination of those states.[1]

The reflectivity spectra for graphene on metal substrates are found to be, overall, similar to those for graphene on SiC. They reveal a series of minima in the energy range 0 – 8 eV, but now it is generally found that *n* layers of graphene produce $n-1$ minima in the reflectivity.[7,8,9] In this result, however, the layer of graphene closest to the substrate is *included* in the count of the number of layers, even though this layer might have electronic properties that deviate from those of graphene due to its bonding to the substrate (we refer to such a layer as *graphene-like*, i.e. with structure similar to that of graphene but with different electronic properties). Such a graphene-like layer also exists for the SiC(0001) surface, known as the "buffer layer",[10] and this graphene-like layer was *not included* in the layer count in the work of Hibino et al.[1] If we *do* include that layer, we then arrive at the result of $n-1$ reflectivity for *n* layers of graphene on SiC, the same as for graphene on metals. Utilization of this revised counting, however, begs the question of how the interface between graphene and the substrate should be properly treated in a full model for the reflectivity spectra.

In addition to this question, there are a number of "irregularities" in the reflectivity spectra that have been noted in recent works. For graphene on SiC(0001), if the graphene-like buffer layer is decoupled from the substrate (e.g. by hydrogenation) then an extra minimum is formed in the spectrum.[3] Depending on the detailed treatment used for the decoupling, this minimum can have a position similar to one of those in the original spectra, or at a higher energy.[5] The same behavior has been reported for graphene on the $SiC(000\bar{1})$ surface, prepared in disilane, for which a graphene-like buffer layer also exists and can be decoupled from the SiC.[6] No theoretical understanding of the energetic locations of these additional minima presently exists.

In this work we develop a theoretical method for computing reflectivity spectra of graphene, and we compare those results with experimentally obtained spectra. For free-standing graphene we demonstrate that *n* layers of graphene actually produce $n-1$ minima in its reflectivity spectrum. The reason that $n-1$ minima are obtained, rather than *n*, is that the wavefunctions for the relevant scattering states are localized *in between* the graphene layers (not *on* them, as in the Hibino et al. model[1]). These states derive from the *interlayer band* of graphite, the structure of which depends sensitively on the exchange-correlation potential in the material.[11,12,13] In our work, we employ a relatively accurate description of that potential, in 3-dimensions, from which we derive the reflectivity of the low-energy electrons. We argue that the pattern of $n-1$ reflectance minima for *n*-layer graphene persists even when the bottommost graphene layer is strongly bonded onto a substrate. However, for a graphene layer that is more weakly bonded onto a substrate we argue that an additional reflectivity is sometimes formed, arising from an interlayer state formed in the space between the graphene and the substrate. The energy of this additional state is typically higher than those of the regular interlayer graphene states, and in this way, the above-mentioned irregularities in the observed spectra can be understood.

For our computations we use the Vienna Ab-Initio Simulation Package (VASP), employing the projector-augmented wave method and the generalized-gradient approximation for the density functional,[14,15,16] with a plane-wave energy cutoff of 500 eV. For graphite, we obtain a band structure which is identical to that displayed by Hibino et al.[1] For free-standing graphene, we simulate the graphene slab surrounded by vacuum of some thickness > 1 nm on either side of the slab. Labeling the direction normal to the slab as $z$, we form linear combinations of the wavefunctions for $k_z > 0$ and $k_z < 0$ such that the waves on one side of the slab have only outgoing character, i.e. an $\exp(+ik_z z)$ transmitted wave. Then, using the same linear combination on the other side of the slab permits us to determine the incident and reflected waves, from which we obtain the reflectivity. Details are provided in the Supplementary Material.

Results are shown in Fig. 1 for the reflectivity spectra of free-standing graphene. We find for an $n$-layer graphene slab (0.335 nm between layers) that there are $n-1$ minima in the reflectivity. The associated wavefunctions are peaked *in between* the graphene layers, as shown in Fig. 2 for the case of 4-layer graphene. These states derive from the image-potential states associated with graphene (all our computation contain two additional eigenvalues slightly *below* the vacuum level associated with symmetric and antisymmetric linear combinations of those states existing on both surfaces of the graphene slab).[17] For the three interlayer spaces displayed in Fig. 2 there are three interlayer states. These interlayer states couple together to form the three transmission resonances seen in the $n=4$ spectrum. Focusing on the real part of the wavefunctions in Fig. 2, the linear combinations are indicated by the labels "+", "0", or "−" on the wavefunction peaks, in accordance with a tight-binding scheme described in detail in Ref. [17].

The computed spectra of Fig. 1 show very good agreement with measured reflectivity curves for multilayer graphene on SiC and other substrates,[1,2,7,9] aside from the occasional presence of higher energy features in those spectra (e.g. with decoupled bottommost graphene layers as discussed in the introductory paragraphs above). However, one significant exception to this agreement occurs for the spectra of graphene on $SiO_2$ reported by Locatelli et al.[18] Those authors report similar results for free-standing graphene and for graphene supported on $SiO_2$. For a single layer of graphene on $SiO_2$ their spectrum displays no strong feature in the reflectivity, in agreement with the $n=1$ case of Fig. 1. However, their 2-layer spectrum displays two reflectivity minima and 3 layers displays three minima, in contradiction to the results of Fig. 1. This significant contradiction calls into question either the experimental or the theoretical results.

Due to this contradiction, we have conducted our own reflectivity measurements of single and multilayer graphene on $SiO_2$ using an Elmitec low-energy electron microscope (LEEM) III. Graphene was first grown on Cu foils by low-pressure chemical vapor deposition (CVD),[19] and then two of these graphene layers were sequentially transferred onto $SiO_2$ covered Si wafers.[20] Samples were cleaned by vacuum annealing for 8 hours at 340°C prior to LEEM study. Experimental electron reflectivity curves for 1 to 4 layers of graphene from these samples are shown in Fig. 3. The corresponding location for each spectrum is indicated in the LEEM image (inset). Identification of the number of layers is made on the basis of the preparation procedure and the resulting film morphology as described in Ref. [20]. For example, the top graphene layer used in this study was non-continuous leaving single-layer regions visible in LEEM images. Three- and four-layer regions come from folds and multilayer nuclei of CVD graphene.

For a single graphene layer we find a somewhat sloping reflectivity, but with no clear minimum. For 2 layers of graphene we find a single reflectivity minimum and for 3 layers we find two minima. We therefore find results which are in good agreement with the theoretical predictions of Fig. 1, at least for $n=2$ and $n=3$ (the energy positions of the minima differ slightly between experiment and theory, but these precise locations involve the separation and interaction between neighboring graphene layers, which could be influenced by residual extrinsic effects in the transferred graphene[21,22]). For the single-layer case the sloping reflectivity is not reproduced in the $n=1$ theory, but this experimental result likely again depends in detail on residual interactions,[21] the corrugation between the substrate and the graphene as further discussed below, and/or the electron transmission of the LEEM due to a particular aperture setting (and, indeed, the spectrum for $n=1$ in Ref. [18] appears much flatter). Our experimental result for $n=2$, with a single well-defined reflectivity minimum, is in disagreement with the prior experimental work of Locatelli et al.[18] However, these same authors in a recent re-examination of their data have identified a spectrum with a single reflectivity minimum,[23] consistent with our interpretation.

The multilayer graphene utilized in our experiments actually consists of *twisted* layers (i.e. without Bernal stacking). Theoretically we expect that this type of twist will produce little change in the reflectivity spectra, since the distance between graphene planes does not change significantly and also the interlayer states that form between the planes have very little (<1%) modulation in their wavefunctions in the directions parallel to the planes. Indeed we find the reflectivity of a twisted bilayer with a $\sqrt{7} \times \sqrt{7}$ - $R38.2°$ structure differs from that of untwisted graphene by less than 0.03 over the entire energy range examined.

We now turn to briefly consider the situation for graphene on a substrate. So long as the bottommost graphene layer that is bonded to the substrate is not so severely distorted as to affect the separation (bonding) between it and the next higher graphene layer, then we expect the spaces between the graphene planes (and the potential therein) will be essentially the same as for free-standing multilayer graphene. Therefore we expect a similar set of interlayer states for the two situations. The only additional consideration is whether or not the space between the bottommost graphene layer and the substrate can itself support an interlayer state. For a relatively small separation $d$ between the bottommost graphene layer and the substrate (i.e. graphene that is strongly bonded to the substrate), we do not expect an interlayer state to form. We have made explicit computations of this situation by considering graphene on Cu(111), using a generalization of the above theoretical methods that will be described elsewhere.[24] We do indeed find that for $d \leq 0.25$ nm the reflectivity spectra are essentially identical for free-standing graphene compared to graphene on the substrate; in both cases there are $n-1$ reflectivity minimum for *n*-layer graphene (including the bottommost graphene-like layer in the count).

For larger $d$ values, i.e. more weakly bound graphene on the substrate, an additional reflectivity minimum occurs in the spectrum, with an energy that decreases as the separation $d$ increases.[24]

For example, at $d = 0.30$ nm this minimum occurs at 9.4 eV whereas for $d = 0.35$ nm it is at 5.1 eV (smaller separations produce confinement of the state, hence shifting it to higher energies). A single layer of graphene on $SiO_2$ might be expected to display this type of reflectivity minimum, but the significant corrugation of the graphene is likely sufficient to inhibit the formation of an interlayer state.[18,25] These numerical results for the energies will vary somewhat depending on the particular substrate, i.e. on the effective potential between the substrate and the bottommost graphene plane. In any case, a qualitative picture for the higher energy "irregular" features that are observed in the experiments emerges from our analysis: an interlayer state can be produced between a weakly bonded (e.g. decoupled) bottommost graphene layer and the substrate. This interlayer state couples to its neighbors, producing an additional reflectivity minimum. Experimentally, this extra minimum is often found to occur at somewhat higher energies compared to the regular series arising from the graphene-graphene separations, presumably because the value of $d$ is somewhat less than a the graphene-graphene spacing of 0.335 nm (and/or the effective potential is higher than that between graphene planes). The situation for graphene decoupled from SiC(0001) by a hydrogenation appears to be a special one in which, coincidentally, the energy of the interlayer state that forms between the decoupled graphene layer and the substrate turns out to be nearly the same as the energy of a graphene-graphene interlayer state.

In summary, we have presented first-principle theoretical results for low-energy reflectivity spectra from free-standing graphene, and compared those to experiment. Good agreement is found (utilizing new experimental results). For $n$-layer graphene, $n-1$ minima occur in the reflectivity spectra over 0 – 8 eV, with these minima being associated with interlayer states that form between the graphene planes. Multilayer graphene with the bottommost layer strongly bonded to a substrate yields a very similar spectra as for the free-standing case. For graphene that is more weakly bonded to a substrate an additional minimum in the reflectivity occur under certain conditions, with an energy that depends on the separation and effective potential between the substrate and the graphene.

Discussions with A. Locatelli, H. Petek, D. A. Stewart, and A. A. Zakharov are gratefully acknowledged. This work was supported by the National Science Foundation and by the Office of Naval Research MURI program. The work at the Naval Research Laboratory was supported by the Office of Naval Research and NRL's Nanoscience Institute. The work at Sandia National Laboratories was supported by the US DOE Office of Basic Energy Sciences (BES), Division of Materials Science and Engineering and by Sandia LDRD. Sandia National Laboratories is a multi-program laboratory managed and operated by Sandia Corporation, a wholly owned subsidiary of Lockheed Martin Corporation, for the U.S. Department of Energy's National Nuclear Security Administration under contract DE-AC04-94AL85000.

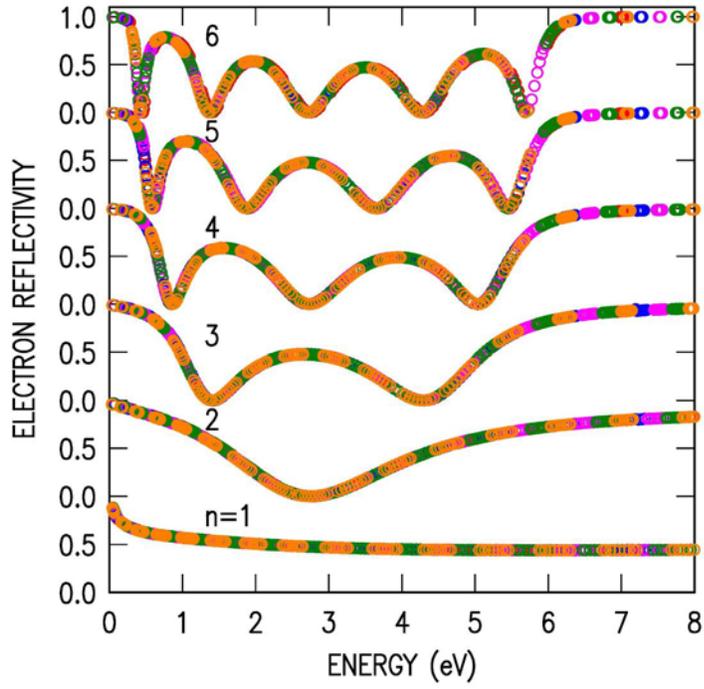

FIG 1. (Color on-line) Computed reflectivity for free-standing slabs of *n*-layer graphene. For each *n*, a series of computation are performed with different vacuum widths; differently shaded (colored) data points are used for plotting the results for each width.

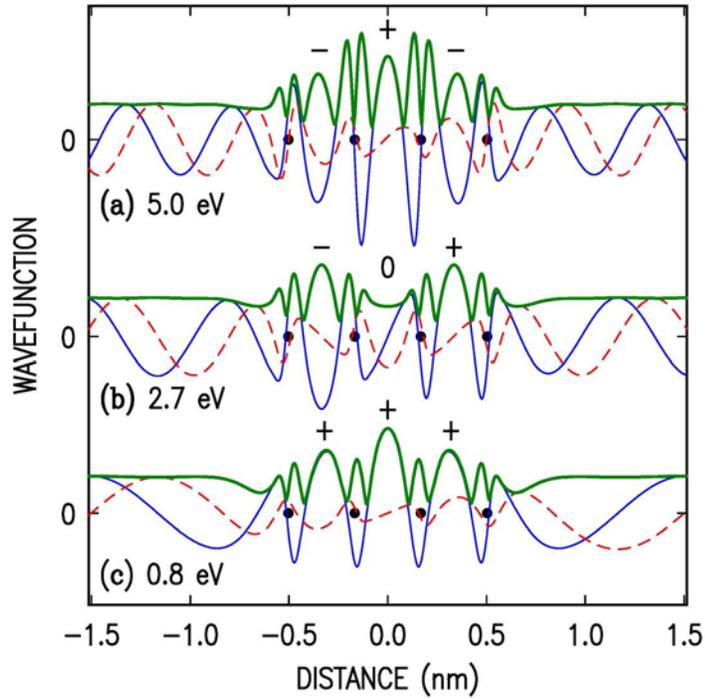

FIG 2. (Color on-line) Wavefunctions at the energies of transmission resonances for $n=4$ layers of free-standing graphene. The real part of the wavefunction is shown by the thin solid line, the imaginary part by the thin dashed line, and the magnitude by the thick solid line (blue, red, and green, respectively, in the color version). The solid black dots indicate the positions of the graphene layers. The +, –, and 0 symbols indicate peaks in the wavefunctions that are concentrated between the graphene layers.

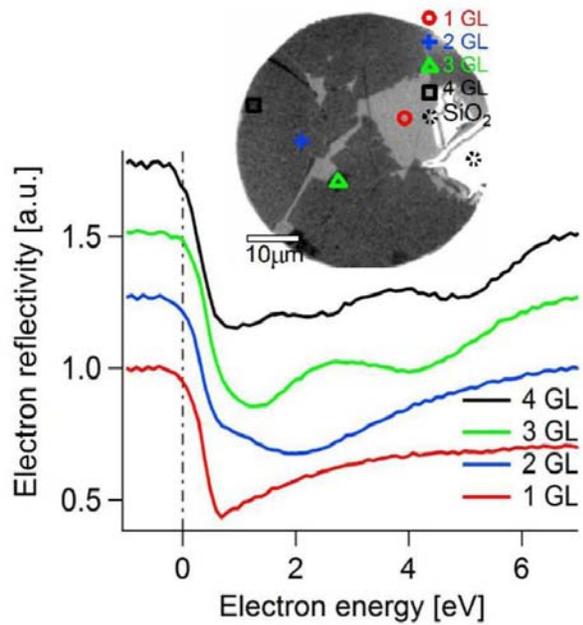

FIG 3. (Color on-line) Measured electron reflectivity for single and multiple graphene layers (GL) on $SiO_2$; curves are vertically offset for clarity. The LEEM image in the inset (acquired at energy 1.8 eV above the sample's vacuum level) shows the surface from where the reflectivity spectra were acquired.

**Supplemental Material**

**Theoretical Method**
Computations are performed in a periodically repeated vacuum-graphene-vacuum system, with at least 1 nm vacuum separating the graphene slabs. Assuming an incoming electron beam at normal incidence, we need only consider states in the graphene having wavevector $\mathbf{k}$ with components $k_x = k_y = 0$ and $k_z \neq 0$ (with $z$ labeling the direction perpendicular to the graphene planes and $z = 0$ being the center of the graphene slab). Two wavefunctions with energy $E_{\nu,k_z}$ are obtained, $\psi_{\nu,k_z}(x, y, z)$ and $\psi_{\nu,-k_z}(x, y, z)$, where $\nu$ is a band index. We work in terms of Fourier components of the wavefunction with reciprocal lattice $\mathbf{G} \equiv (G_x, G_y, G_z)$,

$$\psi_{\nu,\mathbf{k}}(\mathbf{r}) = \sum_{\mathbf{G}} \frac{C_{\nu,\mathbf{G}}}{\sqrt{V}} e^{i(\mathbf{k}+\mathbf{G})\cdot\mathbf{r}} = \sum_{G_x, G_y} \phi_{\nu,k_z}^{G_x,G_y}(z) e^{i[(k_x+G_x)x+(k_y+G_y)y]} \quad (1)$$

where $C_{\nu,\mathbf{G}}$ is a plane-wave expansion coefficient (obtained from the electronic structure computation), $V$ is the volume of a unit cell (included for normalization purposes), and with

$$\phi_{\nu,k_z}^{G_x,G_y}(z) = \sum_{G_z} \frac{C_{\nu,\mathbf{G}}}{\sqrt{V}} e^{i(k_z+G_z)z} . \quad (2)$$

Our evaluations are performed for $\mathbf{k} = (0,0,k_z)$, so we need only consider $\psi_{\nu,k_z}(\mathbf{r})$ in comparison to $\phi_{\nu,k_z}^{G_x,G_y}(z)$. Far out in the vacuum, the wavefunctions $\psi_{\nu,k_z}(\mathbf{r})$ have a specific, separable form: they consist of travelling waves $\exp(i\kappa_{\mathbf{g}} z)$ where $\kappa_{\mathbf{g}}$ labels the $z$-component of the wavevector in the vacuum, multiplied by a sum of lateral waves of the form $A_{\mathbf{g}} \exp[i(g_x x + g_y y)]$ where the lateral wavevector is denoted by $\mathbf{g} = (g_x, g_y)$ and $A_{\mathbf{g}}$ is an amplitude. We have $\kappa_{\mathbf{g}} = \sqrt{2m(E_{\nu,k} - E_V)/\hbar^2 - g_x^2 - g_y^2}$ where $E_V$ is the vacuum energy (corresponding to the potential energy at a $z$-value sufficiently far from the graphene slab so that the potential is essentially constant; we find that a distance of about 0.5 nm is sufficient for this purpose). The lateral wavevector will correspond to one of the $(G_x, G_y)$ values; $(g_x, g_y) = (0,0)$ for the nondiffracted beam and $(g_x, g_y) \neq (0,0)$ for a diffracted beam, the latter existing only for $E_{\nu,k} - E_V \geq \hbar^2(g_x^2 + g_y^2)/2m$. Of course, evanescent states will exist for lower energies, but we are considering distances far enough out in the vacuum so that we do not need to consider those. We note that the values of $\kappa_{\mathbf{g}}$ are quite different than those of $k_z$. For electron energies of 0 – 10 eV the former range over $0 < \kappa_{\mathbf{g}} < 16$ nm$^{-1}$ with $(g_x, g_y) = (0,0)$. The latter are determined by the simulation size in the $z$-direction; with $-z_S < z \leq z_S$ we have $-\pi/2z_S < k_z \leq \pi/2z_S$, which for a typical value $2z_S \approx 5$ nm gives $|k_z| < 0.63$ nm$^{-1}$.

Thus, far out in the vacuum, each eigenstate will consist of a possibly nonzero $\phi_{v,k_z}^{0,0}(z)$ component, together with some number of nonzero $\phi_{v,k_z}^{G_x,G_y}(z)$ components (one for single diffraction or a few for multiple diffraction, along with equivalent components obtained by symmetry operations), where again, the maximum value of $(G_x,G_y)$ is determined by $2m(E_{v,k} - E_V)/\hbar^2 \geq G_x^2 + G_y^2$. In the following discussion we will assume energies low enough so that no diffracted beams occur, which for graphene corresponds to $E < 33.0$ eV with a primitive hexagonal lattice constant of $a = 0.2464$ nm. In this case, any wavefunction that has a significant amplitude far out in the vacuum corresponds to a (0,0) nondiffracted beam. The precise criterion we use to distinguish between a (0,0) beam and other states will be specified shortly. Give this discrimination, we then proceed with the analysis needed to compute the reflectivity.

From the electronic structure computation we obtain $\phi_{v,k_z}^{0,0}(z)$, employing Eq. (2). For our analysis we also require $\phi_{v,-k_z}^{0,0}(z)$, which for the case of a potential that has a mirror plane at $z = 0$ can be easily obtained from $\phi_{v,-k_z}^{0,0}(z) = \phi_{v,k_z}^{0,0}(-z)$. For a nonsymmetric potential we must use Eq. (2) to obtain $\phi_{v,-k_z}^{0,0}(z)$, but in this case we also must ensure that a definite phase relationship exists between $\phi_{v,k_z}^{0,0}(z)$ and $\phi_{v,-k_z}^{0,0}(z)$ (i.e. between $\psi_{v,k_z}$ and $\psi_{v,-k_z}$). This is achieved by taking the phase of $\phi_{v,k_z}^{0,0}(z)$ to be zero at $z = -z_S$ and the phase of $\phi_{v,-k_z}^{0,0}(z)$ to be zero at $z = +z_S$. We then form the linear combinations

$$\phi_{v,+}^{0,0}(z) = [\phi_{v,k_z}^{0,0}(z) + \phi_{v,-k_z}^{0,0}(z)]/\sqrt{2} \qquad 3(a)$$

$$\phi_{v,-}^{0,0}(z) = [\phi_{v,k_z}^{0,0}(z) - \phi_{v,-k_z}^{0,0}(z)]/\sqrt{2} \ . \qquad 3(b)$$

In addition to having a nonzero value for $k_z$, we also perform the evaluations only for $k_z$ not at the edge of the Brillouin zone (i.e. of the simulation cell), thus ensuring that both $\phi_{v,+}^{0,0}$ and $\phi_{v,-}^{0,0}$ are nonzero.

The functions $\phi_{v,+}^{0,0}$ and $\phi_{v,-}^{0,0}$ form standing-wave type states, i.e. any complex phase that they might have is a constant, independent of $z$ [similarly for the linear combinations of the full states $\psi_{v,+} = (\psi_{v,k_z} + \psi_{v,-k_z})/\sqrt{2}$ and $\psi_{v,-} = (\psi_{v,k_z} - \psi_{v,-k_z})/\sqrt{2}$ ]. For a symmetric potential these two functions are even and odd, respectively, although our analysis method works even for asymmetric potentials. Now, consider a range of $z$-values sufficiently far from the graphene slab so that the potential is essentially constant (and equal to the vacuum level); let us label these vacuum regions of $-z_S \leq z \leq -z_L$ on the left-hand side of the slab and $z_R \leq z \leq z_S$ on the right. In these vacuum regions, the standing waves defined by Eqs. (3a) and (3b) can be expressed simply as cosine or sine functions, with some respective phase shifts $\delta_+$, $\delta'_+$, $\delta_-$, and $\delta'_-$, according to

$$\phi_{V,+}^{0,0}(z) = \begin{cases} A_+ \cos(\kappa_0 z + \delta_+), & -z_S \leq z \leq -z_L \quad (4a) \\ A_+ \cos(\kappa_0 z - \delta'_+), & z_R \leq z \leq z_S \quad (4b) \end{cases}$$

$$\phi_{V,-}^{0,0}(z) = \begin{cases} A_- \sin(\kappa_0 z + \delta_-), & -z_S \leq z \leq -z_L \quad (4c) \\ A_- \sin(\kappa_0 z - \delta'_-) & z_R \leq z \leq z_S \quad (4d) \end{cases}$$

where $\kappa_0 \equiv \kappa_{0,0} = \sqrt{2m(E-E_V)}/\hbar$ is the free-electron wavevector. For a symmetric potential we have, of course, $\delta_+ = \delta'_+$ and $\delta_- = \delta'_-$. The phase shifts are determined by searching over the range $-z_S \leq z \leq -z_L$ and $z_R \leq z \leq z_S$ for zeroes of $\phi_{V,-}^{0,0}$ and zeroes in the derivative of $\phi_{V,+}^{0,0}$.

To obtain the reflectivity, we form appropriate linear combinations of $\phi_{V,+}^{0,0}$ and $\phi_{V,-}^{0,0}$ corresponding to only a transmitted wave (traveling in the $+z$ direction) for $z_R \leq z \leq z_S$ and with both incident and reflected waves (traveling in the $+z$ and $-z$ directions, respectively) for $-z_S \leq z \leq -z_L$.[1] To achieve this, we first form the two combinations $\phi_{V,+}^{0,0} \pm i\phi_{V,-}^{0,0}$, which equal $A_+ \cos(\kappa_0 z - \delta'_+) \pm iA_- \sin(\kappa_0 z - \delta'_-)$ far on the right-hand side. We then separate these into their $\exp(+i\kappa_0 z)$ and $\exp(-i\kappa_0 z)$ components and form a further linear combination $A_1(\phi_{V,+}^{0,0} + i\phi_{V,-}^{0,0}) + A_2(\phi_{V,+}^{0,0} - i\phi_{V,-}^{0,0})$ such that the prefactor of the $\exp(-i\kappa_0 z)$ term is zero. This is achieved by taking $A_1 = (A_+ e^{i\delta'_+} + A_- e^{i\delta'_-})/2$ and $A_2 = -(A_+ e^{i\delta'_+} - A_- e^{i\delta'_-})/2$. Forming the same set of linear combinations on the far left-hand side yields a reflectivity (ratio of reflected to incident electron current) of

$$R = \left| \frac{e^{i(\delta'_+ - \delta_-)} - e^{-i(\delta_+ - \delta'_-)}}{e^{i(\delta'_+ + \delta_-)} + e^{i(\delta_+ + \delta'_-)}} \right|^2 \quad (5a)$$

and a transmission of

$$T = \left| \frac{2\cos(\delta'_+ - \delta'_-)}{e^{i(\delta'_+ + \delta_-)} + e^{i(\delta_+ + \delta'_-)}} \right|^2. \quad (5b)$$

For a symmetric potential these formulas simplify to $R = \sin^2(\delta_+ - \delta_-)$ and $T = \cos^2(\delta_+ - \delta_-)$ so that, obviously, $R + T = 1$. From the forms of Eqs. (5a) and (5b) it is not so obvious that $R + T = 1$ for the general case, but we find that this relationship is always satisfied so long as the phase normalization mentioned preceding Eqs. (3a) and (3b) is performed.

The above analysis method is illustrated in Fig. S1. Figure S1(a) shows $\phi_{\nu,k_z}^{0,0}(z)$ for a typical electronic state (this state is the same one discussed in Figs. 1 – 3 of Ref. [2]), and the corresponding $\phi_{\nu,-k_z}^{0,0}(z)$ is pictured in Fig. S1(b). Figures S1(c) and S1(d), respectively, show the resultant $\phi_{\nu,+}^{0,0}$ and $\phi_{\nu,-}^{0,0}$. Figures S1(e) and S1(f) show the linear combinations $\phi_1 \equiv \tilde{\phi}_{\nu,+}^{0,0} + i\tilde{\phi}_{\nu,-}^{0,0}$ and $\phi_2 \equiv \tilde{\phi}_{\nu,+}^{0,0} \pm i\tilde{\phi}_{\nu,-}^{0,0}$, respectively, with $\tilde{\phi}_{\nu,\pm}^{0,0}$ being equal to $\phi_{\nu,\pm}^{0,0}$ multiplied by a phase factor such that the product is purely real. Finally, Fig. S1(g) shows the linear combination $A_1\phi_1 + A_2\phi_2$, where it clear that only an outgoing state is formed on the right-hand side of the slab since the magnitude of the wavefunction is seen to be constant there. Note that this wavefunction at $+z_S$ differs from that at $-z_S$, which is a consequence of that fact that $\phi_{\nu,k_z}^{0,0}(z)$ contains a part that is not periodic over the simulation interval $-z_S < z \leq z_S$ [i.e. a $\exp(ik_z z)$ term].

Returning to the procedure used to test if $\phi_{\nu,+}^{0,0}$ and $\phi_{\nu,-}^{0,0}$ have significant amplitude, we integrate these functions over $-z_S \leq z \leq -z_L$ forming

$$\sigma_\pm \equiv \frac{\sqrt{Az_S}}{z_S - z_L} \int_{-z_S}^{-z_L} \phi_{\nu,\pm}^{0,0}(z) \exp(i\kappa_0 z) \quad \text{and} \tag{6a}$$

$$\sigma \equiv \left[ |\sigma_+|^2 + |\sigma_-|^2 \right]^{1/2} \tag{6b}$$

where $A$ is the area of the unit cell in the $(x,y)$ plane. The $\exp(i\kappa_0 z)$ term occurs in the integrand here since we are considering the inner product of $\phi_{\nu,+}^{0,0}$ and $\phi_{\nu,-}^{0,0}$ with a plane wave propagating in the $+z$ direction. The values for $\sigma$ thus obtained have magnitude near unity for the propagating (0,0) states of interest and are negligible ($\lesssim 10^{-6}$, due to numerical resolution of the computation) for most other states. We can thus set a discriminator for the $\sigma$-values of, say, $10^{-3}$, to include only the states with substantial $\phi_{\nu,k}^{0,0}(z)$ amplitude.

An exception to this sorting of the states is found to occur, when, in certain cases, a state of "mixed" $(g_x, g_y)$ character forms, e.g. a mixed (0,0), (1,0), and (0,−1) state. Such states contain $\phi_{\nu,k_z}^{1,0}(z)$ and $\phi_{\nu,k_z}^{-1,0}(z)$ terms that dominate the wavefunction within the graphene slab, but they also contains a small, nonzero $\phi_{\nu,k_z}^{0,0}(z)$ term that constitutes the only contribution to the wavefunction far out in the vacuum. Such states appear to be precursors to diffracted states, i.e. at higher energy the $\phi_{\nu,k_z}^{1,0}(z)$ and $\phi_{\nu,k_z}^{-1,0}(z)$ terms would have nonzero amplitude in that vacuum and they could be combined to form diffracted beams. For these mixed states, the measure of σ produces values intermediate between $10^{-6}$ and 1, with values in the $10^{-2}$ or $10^{-1}$ range specifically found to occur. Again, it is necessary to reject these mixed states from the analysis, and for this reason we typically use a discriminator for the σ-values that is close to 1; a value of 0.8 is used for all results reported in this work and this value works for the present computations to reject all mixed states.

Even though our analysis method for the (0,0) beam relies on the use $\phi_{\nu,k}^{0,0}(z)$, i.e. the (0,0) Fourier component of the wavefunction, it should be emphasized that the method does indeed fully include all multiple scattering within the slab. The electronic structure solutions do, of course, contain all of that multiple scattering, and we fully employ those solutions in our analysis. Our use of the (0,0) Fourier component is made, in essence, in order to match the full wavefunction to a plane wave far out in the vacuum.

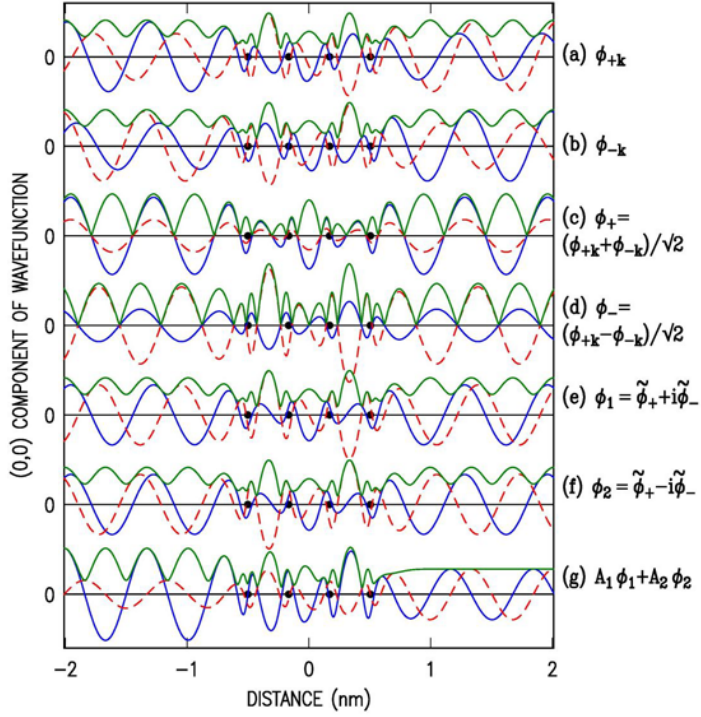

FIG S1. Wavefunctions for various states of 4-layer graphene, plotted over the entire simulation interval $-z_S < z \leq z_S$ and illustrating the process used to deduce the reflectivity: (a) a typical state, having wavevector $k_z = 0.39$ nm$^{-1}$ and energy 3.24 eV above the vacuum level, (b) the corresponding state with negative wavevector, (c) and (d) linear combinations of these two $+k_z$ and $-k_z$ states. (e) and (f) Further linear combinations, which ultimately yield the state shown in (g) that has only an outgoing plane wave on the right-hand side of the graphene slab (see text). The real-part of the wavefunctions are shown by a solid blue line and the imaginary parts by a dashed red line, with the magnitude shown by a solid green line.

**References**

---

[1] The reader is reminded that $\phi_{V,+}^{0,0}$ and $\phi_{V,-}^{0,0}$ are equal to $\psi_{V,+} = (\psi_{V,k} + \psi_{V,-k})/\sqrt{2}$ and $\psi_{V,-} = (\psi_{V,k} - \psi_{V,-k})/\sqrt{2}$, respectively, far out in the vacuum, so this linear combination procedure yields the appropriate *full wavefunctions* of the incident, transmitted and reflected waves. Actually, the entire analysis procedure can be performed using $\psi_{V,+}(x,y,z)$ and $\psi_{V,-}(x,y,z)$ rather than $\phi_{V,+}(z)$ and $\phi_{V,-}(z)$. The only difference is that, in the former case, a slight dependence might exist in the results on the *x* and *y* locations used for the evaluations, i.e. if the distance out into the vacuum is insufficiently large. This dependence is entirely absent when using $\phi_{V,+}(z)$ and $\phi_{V,-}(z)$.

[2] R. M. Feenstra and M. Widom, arXiv:1212.5506 [cond-mat.mes-hall] (2012), submitted to Ultramicroscopy.